\documentstyle[prb,twocolumn,aps]{revtex}
\tighten
\begin{document}

\title{Non-linear Effects in Resonant Tunnelling  Through a Quantum Dot}
\author{E. S. Rodrigues$^1$, E.V.Anda $^2$ and P. Orellana $^3$}
\address{$^1${\it Instituto de F\'{\i}sica, Universidade Federal Fluminense,}\\
{\it Gragoat\'a, Niter\'oi RJ,Brasil, 24210-340.},\\
$^2${\it Departamento de F\'{\i}sica, Pontificia Universidade Cat\'olica de Rio do Janeiro,}\\
{\it Caixa Postal 38071, Rio de Janeiro, Brasil, 22452-970.}, \\ 
$^3${\it Departamento de F\'{\i}sica, Universidad Cat\'olica del Norte,}\\
{\it Av. Angamos 0610, Casilla 1280, Antofagasta, Chile.}}
\address{}
\address{\mbox{ }}
\address{\parbox{14cm}{\rm \mbox{ }\mbox{ }\mbox{ }
We study resonant tunnelling transport properties of a quantum dot
connected to two leads. The flowing electrons are supposed to
interact strongly at the quantum dot. The system is represented by an
Anderson impurity Hamiltonian. The transport properties are
characterised by the well-known Coulomb blockade properties. However,
superimposed to them, this {\bf 1D} system possesses novel non-linear
current bistability behaviour and other instability phenomena,
which leads eventually to chaos.
}}
\address{\mbox{ }}
\address{\parbox{14cm}{\rm PACS numbers:  73.40.Gk,73.40.-c,72.15.Gd}}

\maketitle

\makeatletter

\global\@specialpagefalse
\def\@oddhead{\underline{To appear in Solid State Communication.\hspace{281pt}
}}
\let\@evenhead\@oddhead
\makeatother

\narrowtext

\section*{Introduction}
Transport properties have been extensively studied due to their
interesting device application and very rich phenomenology. Since
the first observation of resonant tunnelling in a {\bf 3D} double barrier
structure by Chang, Esaki and Tsu~\cite{ll}, these devices have been found
to present non-linear properties, which are reflected in the
observation of multistabilities in the I-V characteristic curve.
Besides a peak due to simple resonant transmission, dynamical
intrinsic bistability and hysteresis in the negative differential
resistance region has been measured~\cite{vj,az}. These properties have 
been explained as a non-linear effect due to the Coulomb interactions
among the flowing charges~\cite{az}. The phenomenon can be thought to be
produced by the rapid leakage of the electronic charge accumulated at
the well between the barriers when the applied potential is just
taking the device out of resonance~\cite{pp}. Very recently we were able to
show that the addition of a magnetic field parallel to the current
induces self-sustained intrinsic current oscillations in an
asymmetric double barrier structure~\cite{po}. These oscillations were
attributed to the enhancement due to the external magnetic field of
the non-linear dynamic coupling of the current to the charge trapped
in the well. These results showed that the system bifurcates as the
field is increased, and may transit to chaos at large enough fields.
Although the presence of the external magnetic field introduces some
{\bf 1D} character to these systems, all the properties are obtained from {\bf
3D}
systems.  In recent years, electron channels have been fabricated with
a lateral quantum confine energy levels sufficiently separated one
from the other as to be able to study properties of a {\bf 1D} system. When
a quantum dot is connected through these electron channels to a
battery, the system exhibits pronounced periodic oscillations of the
conductance when the state of charge of the dot is modified by
changing the gate potential. This is a consequence of the Coulomb
blockade of an electron when it tries to go into a quantum island,
which is already occupied by another electron~\cite{jh,fh}. Although the
multistability phenomena found in standard {\bf 3D} resonant tunnelling
devices have not been detected in {\bf 1D} systems where the Coulomb
blockade is dominant, it is evident that both phenomena derive from
the same Coulomb interaction acting between the flowing charges.  In
this paper we developed a self-consistent theory of resonant
tunnelling in quasi-{\bf 1D} systems which predicts that, besides the normal
Coulomb blockade behaviour in the region of negative differential
resistance, the system develops multistable properties. Moreover we
show that, while the phenomena of hysteresis persists, the system
bifurcates and is capable of further bifurcations as the gate
potential applied to the dot is increased, leading eventually to true
chaos.  This fact emphasizes the absence of stationary solutions for
the current in certain regions of the parameter space, indicating the
existence of time-dependent phenomena, probably oscillations.

\section*{The Model}
A theoretical understanding of these properties should consider that
the system is in a non-equilibrium situation under the effect of
many-body interactions. A {\bf 1D} Anderson Hamiltonian microscopically
describes the system where the quantum dot plays the role of the
impurity. The leads are represented by a {\bf 1D} tight-binding Hamiltonian
connecting the dot to two reservoirs characterised by Fermi levels
${\cal E}_l$ and ${\cal E}_r$. The difference ${\cal E}_l-{\cal E}_r$ corresponds to the
potential drop from left to right along the sample.  The
nearest-neighbour Hamiltonian can be written as,

\begin{eqnarray}
{\cal H} &=& t\sum_{<ij> \sigma}c^{\dagger}_{i\sigma}c_{j\sigma}+
\sum_{i \sigma} 
\epsilon_i n_{i \sigma} + {\cal V}_p n_{o\sigma} + {{\cal U} \over 2}
\sum_{\sigma} n_{o\sigma}
n_{o{\overline{\sigma}}} \nonumber \\
& & + \hskip .1cm t_r\biggl(c^{\dagger}_{o\sigma} c_{1\sigma} +
c.c.\biggr) + t_l\biggl(c^{\dagger}_{o\sigma}
c_{{\overline{1}}\sigma} + c.c.\biggr)
\label{1}
\end{eqnarray}
\noindent where ${\cal V}_p$ is the external gate which controls the
energy of the localised state within the well, ${\cal U}$ is the
local electronic repulsion of the charges inside the well localized
at site $0$, t is the nearest-neighbour hopping matrix elements,
$t_r$ and $t_l$ are the coupling matrix elements between the well
and the right and left contacts. It is assumed the system has $N$
sites. The electron-electron interaction will be treated within the
context of the Hubbard approximation~\cite{jh}, which
is adequate to treat the Coulomb blockade and the non-linear
effects derived from the Coulomb interaction.  However, it will not
be capable of describing the low-lying excitations associated to the
Kondo effect which in principle the system would have at temperatures
below the Kondo temperature. On a tight-binding basis, the stationary
state of energy ${\cal E}_k$ can be written as
\begin{equation}
\varphi_{k \sigma} = \sum_i a_{i \sigma}^k \phi_{i \sigma}(r)
\end{equation}
\noindent where $\phi_{i \sigma}$ is a Wannier state localized at
site $i$ of spin $\sigma$ and the coefficients $a_{i \sigma}^k$ obey
the non-linear difference equations,
\begin{equation}
\biggl(\epsilon_i - {\cal E}_k + 2t\biggr) a^k_{i,\sigma} = 
\biggl(a^k_{i-1,\sigma} + a^k_{i+1,\sigma}\biggr)t \hskip 2.9cm 
\mbox{; }i\ne 0
\label{2}
\end{equation}
$$\biggl(\epsilon_o - {\cal E}_k + 2t + {\cal V}_p +
{\cal U}\biggr) a^k_{o,\sigma} =$$
\begin{eqnarray}
a^k_{-1,\sigma}\biggl(t + \frac{{\cal
U}t_{l}n^-_{\sigma}}{-{\cal E}_k+2t+{\cal V}_p}\biggr) +
a^k_{1,\sigma}\biggl(t + \frac{{\cal U}t_{r} n^-_{\sigma}}{-{\cal
E}_k + 2t + {\cal V}_p}\biggr) \hskip .5cm \mbox{; }i=0
\label{3}
\end{eqnarray}
\noindent where $n^{-}_{\sigma} = 1 - <n_{o,\sigma}>$ and
$<n_{o,\sigma}>$ is the number of electrons with spin $\sigma$ at the
well site and is calculated from the tight-binding amplitudes at the
well, according to the equation,
\begin{equation}
<n_{o,\sigma}> = \sum_{k<k_f^{l}} \left\vert a_{o \sigma}^k
\right\vert^2
\end{equation} 
\noindent where the sum over $k$ covers all occupied electrons states of the
system with energy below the Fermi level, incident from the emitter
side.  In order to study the solutions of equation [2] we assume a
plane wave incident from the left with an  intensity $I$, with a
partial reflection amplitude $R$. The waveform at the far right is a simple plane
wave with 
intensity given by the transmission probability $T$. Taking this to be
the solution at the far right and left edges of the system, we can
write,
\begin{equation}
\varphi^k (r) = I e^{ikr} + R e^{-ikr} \hskip 1cm r \ll 0
\end{equation}
\begin{equation}
\varphi^k (r) = T e^{ikr} \hskip 2.6cm r \gg L
\end{equation}
\noindent where $L$ is the length of the active part of the system in
units of the lattice parameter. The solution of equations [3,4] can be
obtained through an adequate iteration of it from right to left.
For a given transmitted amplitude, the associated reflected and
incident amplitudes may be determined by matching the iterated
function to the proper plane wave at the far left, equation [6]. The
transmission square modulus of T, adequately normalised by the incident
amplitude I, obtained from the iterative procedure, multiplied by  the
wave vector k, gives us the contribution of this wave vector to the
current.  Due to the presence of the non-linear term, the equations [3,4]
have to be solved self-consistently. With this purpose, we define a
second pseudo-time-like iteration in the following way. Initially Eqs.
[3,4] are solved ignoring the non-linear term and for energies up to the
Fermi energy. The coefficients thus obtained correspond to a solution
for non- interacting electrons. They are used to construct the
non-linear term for the next solution. The procedure is continued,
using for the non-linear term solutions corresponding to the previous
iteration.  This procedure defines a spatial-pseudo-temporal map that
is linear in space and non-linear in time. Maps non-linear in space
and time have been studied in other contexts. In particular, the
coupled map model for open flow has been found to exhibit spatial
chaos with temporal periodicity~\cite{fh}. A similar behaviour has been
found studying resonant tunnelling properties of a {\bf 3D} double barrier
devices under the effect of an external magnetic field along the
current (4). As we shall show, in our case chaos develops in the
iteration of time, as the self-consistent loop is developed.

\section*{Numerical Results}
We study a model which consists of leads connected to a quantum dot
represented by $t_r =0.1t$, $t_l= 0.1t$, ${\cal U}=1.0t$ and a Fermi
level $e_f = 0.05t$. These parameters could adequately represent a
semiconductor device constructed from a GaAs substrate. The
normalisation of the wave function is taken so that each site could
have up to a maximum of two electrons. The map is linear in space so
that the normalisation involves a simple multiplication of all
coefficients by a constant.  Since the electron density has rather
long-range oscillations, we made sure the sample was long enough to
make finite size effects negligible. This is guaranteed by taking  a
sample of $N>100$ sites.

Fig 1 shows the current voltage (I-V) characteristics for
several values of the gate potential ${\cal V}_p$ applied at the quantum dot.  
As it is shown in the figure for this {\bf 1D} system, we obtain a similar
behavior to the very well-known bistability in {\bf 3D} systems. There are two
solutions for the current as the voltage is increased when
the dot is charged (solid line), and as the voltage is decreased
when the dot is without charge (stars). In either
case, the self-consistent solution converges to a stable fixed point
after some finite number of iterations. Similarly to {\bf 3D} tunnelling, 
the choice of an asymmetric structure 
($t_r > t_f$) and an increase of the Fermi level augments the charge
content of the well and consequently the non-linearities and in
particular the width of the hysteresis loop for this {\bf 1D} system.  
When the gate 
potential is greater than a threshold value, which depends upon the
parameters of the system, a completely novel feature starts to
develop where two periods and two fixed points are encountered for
certain voltages as shown in Fig 1b.  This bubble-like solution is
obtained as the voltage is increased outside the region of
bistability showing that although both phenomena are derived from the
non-linearities introduced by the local Coulomb repulsion, they
correspond to two different effects.  Note that these new solutions
are not reached, as the voltage is decreased (stars) as it
corresponds to a discharged dot where the non-linearities are
negligible. As the potential gate is raised still further, the area
of the bistable bubble increases and it enters into the bistable
region. At larger gate potentials the solution undergoes further
bifurcations and finally a bifurcation cascade leading to a chaotic
region as it is shown in fig1c and 1d for a gate potential ${\cal V}_p = -0.7t$
and $-1.4t$ respectively. 

As the external applied potential increases still further, the chaotic
region eventually unfolds to a solution of finite periodicity or to a
unique self-consistent one.  In fig.2 a phase diagram in the
parameter space $V-{\cal V}_p$ shows the various structures we have
obtained. The thick lines define the region of bistability, while the
stars delimit the region where bifurcations of various orders and chaotic behaviour take place.

\section*{Conclusions}
We have shown that a quantum dot connected to two leads and under the
effect on an external applied potential have, superimposed to the
Coulomb blockade oscillations of the current, bistable solutions and
instabilities which eventually become chaotic. Although bistabilities in the
current are a well-known phenomenon in {\bf 3D} resonant tunneling, to the best of 
our knowledge, neither its existance has been proposed nor  has it been measured in a {\bf 1D} system. Bistabilities and general instabilities, 
although arising from the same
Coulomb interaction, appear in different regions of the parameter space
reflecting the fact that they are different phenomena.  A full-time
dependent study of this problem is under way.

\section*{Acknowledgments}
This work was partially supported by the Brazilian Agencies CNPQ and Finep, 
DGICT-UCN, FONDECYT Grants 1980225 and 1960417 (Chile).

\begin{figure}[h]
\vspace{2.0 cm}
\caption{
$I-V$ characteristic curve for a system with ${\cal U} = 1t$; $t_r = 
t_l = 0.1t$ for four different values of the gate potential; a) ${\cal V}_p = 
-0.1t$, b) ${\cal V}_p = -0.2t$, c) ${\cal V}_p = -0.7t$, d) ${\cal V}_p = -1.4t$.}
\label{fig1}
\end{figure}

\begin{figure}[h]
\caption{
Phase diagram in the parameter space $V-{\cal V}_p$ for the 
system of Fig1. The thick lines and stars  delimit the region of
bistability 
and instability respectively.} 
\label{fig2}
\end{figure}
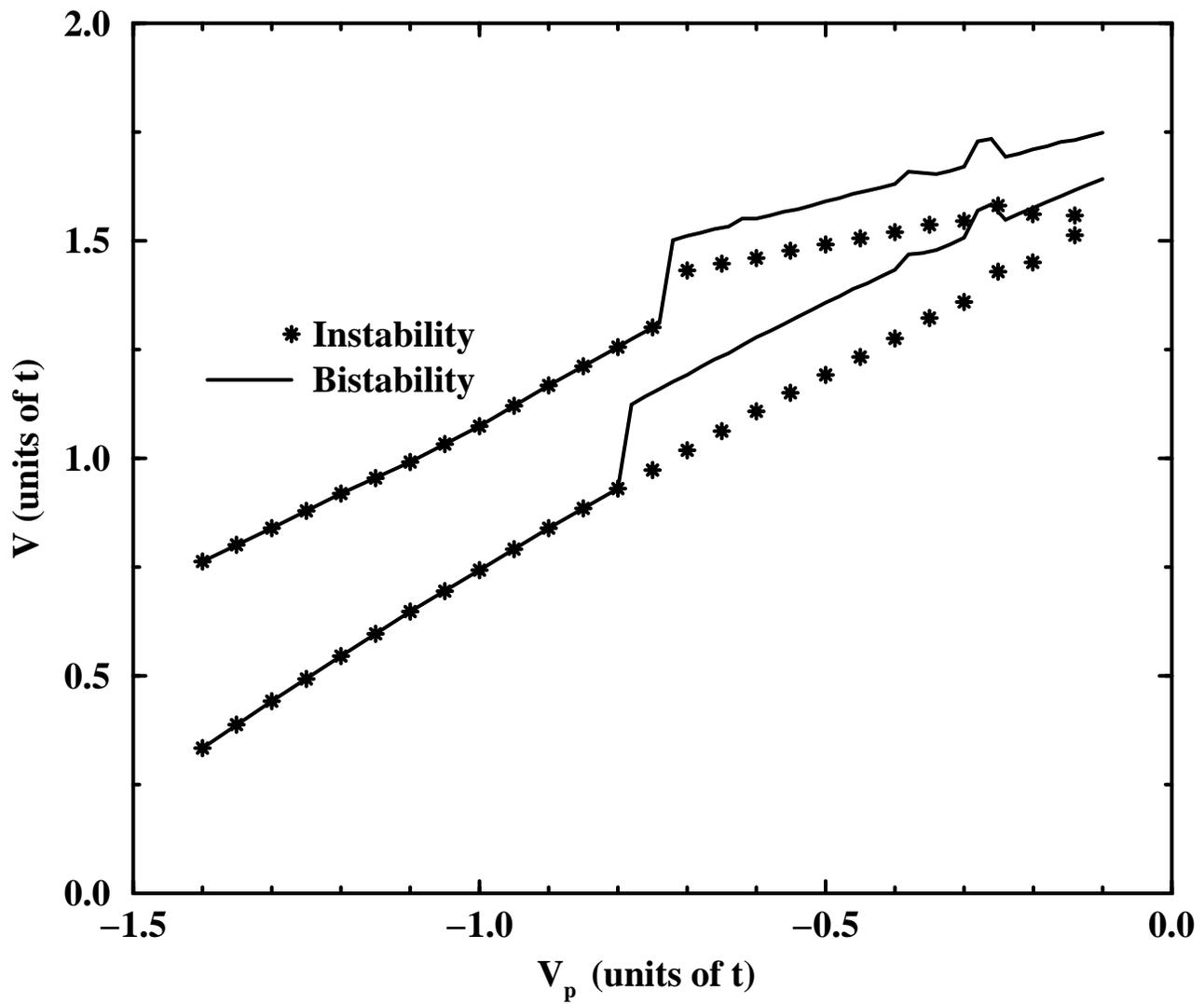
\end{document}